\documentclass[
prl,twocolumn,
amsfonts,amsmath,amssymb, 
showpacs,
superscriptaddress]{revtex4}
\usepackage{graphicx}
\usepackage[latin1]{inputenc}
\usepackage[T1]{fontenc}
\begin{document}
\title{Nonlinear mode-coupling and synchronization of a vacuum-trapped nanoparticle}

 \author{
 Jan Gieseler
 }
 \affiliation{ICFO-Institut de Ciencies Fotoniques, Mediterranean Technology Park,
 08860 Castelldefels (Barcelona), Spain}

\author{
Marko Spasenovi\'c
}
 \affiliation{ICFO-Institut de Ciencies Fotoniques, Mediterranean Technology Park,
 08860 Castelldefels (Barcelona), Spain}
 \affiliation{Institut of Physics, University of Belgrade, Pregrevica 118, 11080 Belgrade, Serbia}

\author{
Lukas Novotny
}
 \affiliation{Photonics Laboratory, ETH Z\"urich, 8093 Z\"urich, Switzerland} 

\author{
Romain Quidant
}
 \affiliation{ICFO-Institut de Ciencies Fotoniques, Mediterranean Technology Park,
 08860 Castelldefels (Barcelona), Spain}
 \affiliation{ICREA-Instituci{\'o} Catalana de Recerca i Estudis Avan\c{c}ats, 08010 Barcelona, Spain}

\date{\today}

\begin{abstract}
We study the dynamics of a laser-trapped nanoparticle in high vacuum. Using parametric coupling to an external excitation source, the linewidth of the nanoparticle's oscillation can be reduced by three orders of magnitude.
We show that the oscillation of the nanoparticle and the excitation source are synchronized, exhibiting a well-defined phase relationship. Furthermore, the external source can be used to controllably drive the nanoparticle into the nonlinear regime, thereby generating strong coupling between the different translational modes of the nanoparticle. Our work contributes to the understanding of the nonlinear dynamics of levitated nanoparticles in high vacuum and paves the way for studies of  pattern formation, chaos, and stochastic resonance. 
\end{abstract}

\maketitle

Synchronization of spatially separate processes occur in biological, chemical, physical, and social systems, and have attracted the interest of scientists for centuries.
Nanomechanical oscillators naturally lend themselves to the experimental \cite{Eichler:2012ju,Villanueva:2013cv,Karabalin:2009dt,Lulla:2012gq,Matheny:2013cm} and theoretical \cite{Lifshitz:2008td} study of nonlinear behavior and synchronization.
The nonlinear regime can be exploited for applications including phonon-cavity cooling \cite{Mahboob:2012cl,Venstra:2011ij}, precision frequency measurements \citep{Aldridge:2005hx}, signal amplification via stochastic resonance \citep{Almog:2007cn, Westra:2013du}, mass sensing \citep{Buks:2006kk} and quantum non-demolition measurements \cite{Caves:1980ww,Braginsky:1980hj,Unruh:1978gx}. 
In addition, nonlinear mechanical oscillators have been proposed as memory elements \cite{Mahboob:2008dk,Bagheri:2011cr} and to hallmark classical to quantum transitions \cite{Katz:2007kg}.\\[-1ex]

Recently, optically levitated nano- and microparticles \cite{Gieseler:2012bi,Raizen:2012gn,Li:2011jl,Kiesel:2013bp} have raised great interest because of their exceptional mechanical properties. Unlike other nanomechanical oscillators, both the frequency and the $Q$-factor of a levitated particle can be precisely controlled in situ by adjusting the laser power and the gas pressure, respectively.
Yet, due to difficulties in reaching high vacuum with levitated nanoparticles \cite{Kiesel:2013bp,Monteiro:2013bx}, there have only been  few experimental studies \cite{Gieseler:2012bi,Gieseler:2013ip} of oscillator dynamics in high vacuum.
In this Letter, we investigate the response of a levitated nanoparticle in high vacuum to single frequency excitations.
In addition to nonlinear coupling between translational modes and synchronization with an external source, we identify collisions of the driven particle with residual air molecules as additional features in the spectrum of the particle motion.\\[-1ex]

\begin{figure}[!b]
 \begin{center}
\includegraphics[width=0.45\textwidth]{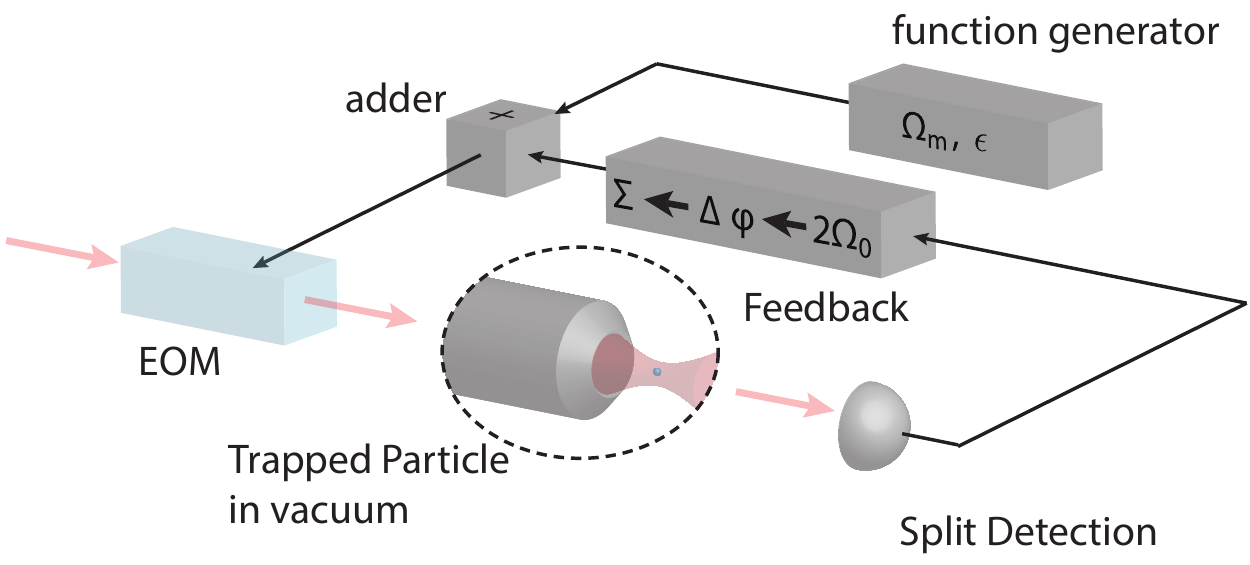}
\end{center}
\caption{
Experimental configuration.
A $\rm SiO_2$ nanoparticle ($a\sim 75 \rm nm$) is trapped by a tightly focused laser beam.
The translational degrees of freedom of the nanoparticle are measured with photodetectors and
the center-of mass motion is cooled down by  parametric feedback  \cite{Gieseler:2012bi}. In addition to feedback, we apply a resonant parametric modulation to excite the particle into the nonlinear regime.
\label{fig:ExperimentalSetup}}
\end{figure}

The experimental configuration is shown in Fig.~\ref{fig:ExperimentalSetup}.
A $\rm SiO_2$ nanoparticle of radius $a\sim 75 \rm nm$ is trapped at the focus of a single beam optical tweezer \cite{Gieseler:2012bi}.
The motion of the particle is imprinted on the phase of the light scattered by the particle.
In the forward direction, the scattered light interferes with the transmitted beam giving rise to an intensity pattern that depends on the particle position.
Using a balanced split detection scheme, the intensity pattern yields information about the three dimensional particle position in real time with a noise floor of $1 \rm pm/\sqrt{ Hz}$ \cite{Gieseler:2012bi}.\\[-1ex]

The particle motion consists of three modes, each corresponding to a spatial oscillation along one of the three symmetry axes of the optical intensity distribution.
The gradient of the optical intensity distribution exerts a restoring force $F_i^{\rm grad}=-k_i \left(1+\sum_{j=x,y,z} \xi_{j} x_j^2\right)x_i$ on a dipolar particle that is displaced from the trap center by $x_i$.
Here, the linear trap stiffness is given by $k_i=\alpha E_0^2/{\rm w}_i^2$, where $E_0$ is the electric field intensity at the focus, $\alpha$ is the polarizability, and ${\rm w}_i$ is the beam waist radius  ($x$,$y$) or Rayleigh range ($z$), and the nonlinear coefficients are given by $\xi_j=-2/{\rm w}_j^2$ \cite{Gieseler:2013ip}.
The nonlinear behaviour and intermode coupling can be understood in terms of a Gaussian model of the focal intensity distribution \cite{Gieseler:2013ip}.
The focal intensity distribution becomes wider as the distance to the focus increases and consequently, the restoring force $F^{\rm grad}_i$ becomes weaker as the motional amplitude increases.
For a sphere of radius $a$ and dielectric constant $\epsilon_p$, the quasi-static polarizability is  $\alpha=4\pi a^3\epsilon_0(\epsilon_p-1)\left/(\epsilon_p+2)\right.$, $\epsilon_0$ being the vacuum permittivity.
Due to the asymmetry of the optical focus, the oscillation frequencies $\Omega_i=(k_i/m)^{1/2}$ along the three main axes are different ($\Omega_z/2\pi\sim37\rm kHz,\;\Omega_x/2\pi\sim125\rm kHz,\;\Omega_y/2\pi\sim135\rm kHz$), $m$ being the mass of the particle.
By means of parametric feedback cooling we are able to reach pressures of $0.5\times 10^{-6}\rm mBar$, where we measure $Q$-factors of $2\times 10^8$ \cite{Gieseler:2012bi,Gieseler:2013ip}.
Under the action of feedback cooling, the effective thermal amplitude $q_{\rm eff}=\left(2k_BT_{\rm eff}\left/m \Omega_i^{ 2}\right.\right)^{1/2}$ of the particle oscillation is kept much smaller than the size of the trap.
As a consequence, coupling between the modes is negligible in the absence of external modulation. \\[-1ex]

However, in addition to parametric feedback cooling we also apply parametric modulation of frequency $\Omega_{\rm m}\approx 2\Omega_i$ with modulation depth $\epsilon$. This modulation acts predominantly on the mode $\Omega_i$ (Fig. \ref{fig:ExperimentalSetup}) and allows us to control its oscillation amplitude. For large oscillation amplitudes the oscillator is driven into its nonlinear regime, where the modes couple through cubic nonlinearities in the optical force. Taking the three terms together (cooling, modulation, and nonlinearities)  the particle's equation of motion becomes \cite{Gieseler:2013ip}
\begin{align}\label{eq:EoMDuffing}\nonumber
&\Omega_i^2 \left[1+\underbrace{\epsilon\cos\left(\Omega_{\rm m}t\right)}_{\rm parametric\; drive}+\underbrace{\Omega_i^{-1}\eta q\dot{q}}_{\rm feedback}+\underbrace{\xi q^2}_{\rm Duffing\; term}\right]q\\[-1ex]
&+\ddot{q} +\Gamma_0 \dot{q}\;=\;\frac{\mathcal{F}_{\rm fluct}}{m }\;\approx \;0.
\end{align}
Here, $\eta$ is the nonlinear damping due to feedback cooling and $\Gamma_0$ is the (linear) damping due to collisions with residual air molecules. It is related to the stochastic force by the fluctuation-dissipation theorem $\langle \mathcal{F}_{\rm fluct}(t)\,  \mathcal{F}_{\rm fluct}(t')\rangle = 2 m  \Gamma_0 \, k_B  T_0 \;\!\delta(t-t')$, $k_B$ and $T_0$ being the Boltzmann constant and the temperature of the environment, respectively.  
Since the three terms in the bracket (parametric driving, feedback, Duffing term) are much stronger than the stochastic force $\mathcal{F}_{\rm fluct}$, the latter can be neglected and the problem reduces to solving a deterministic equation of motion.\\[-1ex]

For the following we consider the particle dynamics at low pressures ($Q\gg 1$)
where the change in the oscillation amplitude is slow compared to the oscillation frequency.
Thus, the solutions $q(t)$ can be described by
 the ansatz
\begin{equation}\label{eqn:Ansatz}
  q(t)=\frac{q_0}{2} A(\tau)\, e^{i\Omega_m t/2}+c.c.,
\end{equation}
where we have introduced the dimensionless slow time scale $\tau=\kappa \Omega_i t$ and the slowly varying displacement amplitude $A(\tau)$ with scale factors $\kappa=\Gamma_0/\Omega_i=Q^{-1}$ and $q_0^2=\kappa/\xi$.
With ansatz \eqref{eqn:Ansatz} we obtain an equation of motion for
 $A(\tau)$ \cite{Lifshitz:2008td}
\begin{equation}\label{eqn:EoMforA2}
 \frac{{\rm d}A}{{\rm d}\tau}=- \frac{\tilde{\gamma}_0}{2}A+i\frac{\tilde{\delta}_{\rm m}}{2}A-\left[\frac{1}{8}\tilde{\eta}-i\frac{3}{8}\right]|A|^2A+i\frac{\tilde{\epsilon}}{4}A^*,
\end{equation}
where $\tilde{\delta}_{\rm m}=\delta_m/\kappa$ is the rescaled normalised detuning $\delta_m=(2-\Omega_{\rm m}/\Omega_i)$. We have also introduced the normalized parameters  
$
\tilde{\gamma}_0=\Gamma_0\left/\Omega_i\kappa\right.,\;
\tilde{\eta}=\eta\left/\xi\right.,{\rm and }\;
\tilde{\epsilon}=\epsilon\left/\kappa\right..
$
Equation \eqref{eqn:EoMforA2} has up to three steady state solutions $A_0(\tau)$, which fullfil ${\rm d}A_0/{\rm d}\tau=0$.
The existence and stability of the steady state solutions depend on the modulation parameters $\epsilon$ and $\delta_{\rm m}$.
Linearization around the steady state yields one unstable solution and two stable solutions.
The first stable solution is the trivial low amplitude solution $A_0=0$.
It is stable if the linear stability condition 
\begin{equation}\label{eq:Lock-inStabilityCondition}
\epsilon<\frac{2}{Q}\sqrt{1+Q^2\delta_{\rm m}^2}\approx 2\delta_{\rm m}
\end{equation}
is fulfilled.
The second stable solution is the high amplitude solution \cite{Lifshitz:2008td}
\begin{align}\label{eqn:ao2}
  q^2&=q_0^2|A_0|^2=
  -\frac{1}{\eta\delta_{\rm th}^2}\\[-1ex]
  &\times\left[3\frac{\xi}{\eta}\delta_m+Q^{-1}+\sqrt{\delta_{\rm th}^2\epsilon^2-\left(\delta_{\rm m}-3\frac{\xi}{\eta}Q^{-1}\right)^2}\right]\nonumber \\[-1ex]
 & \approx -\frac{1}{\eta\delta_{\rm th}^2}\left[3\frac{\xi}{\eta}\delta_m+\sqrt{\delta_{\rm th}^2\epsilon^2-\delta_{\rm m}^2}\right],\nonumber 
\end{align}
which is stable if the nonlinear stability condition
\begin{equation}\label{eq:Lock-inStabilityConditionNL}
\epsilon>|\frac{\delta_m}{\delta_{\rm th}}|,
\end{equation}
with $\delta_{\rm th}=\sqrt{9\xi^2+\eta^2}\left/2\eta\right.$, is fulfilled.
In \eqref{eqn:ao2} we have converted back to physical quantities to facilitate the interpretation of the experimental results. The approximation used in (\ref{eqn:ao2}) holds for $Q\gg 1$, which is the case for typical experimental parameters.\\[-1ex]

\begin{figure*}[!t]
 \begin{center}
\includegraphics[width=0.9\textwidth]{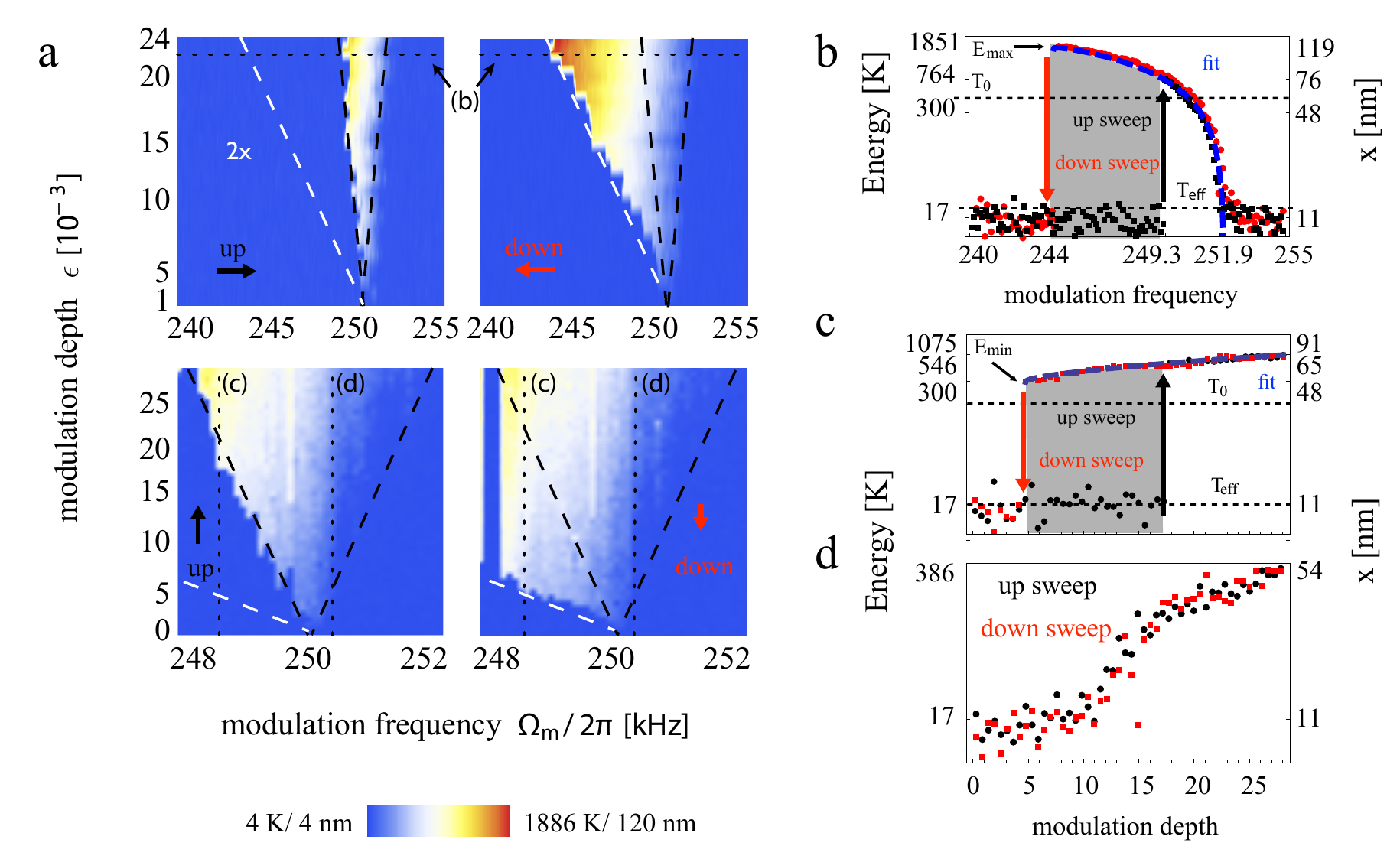}
\end{center}
\caption{
Particle response to parametric modulation.
(a) In the $\epsilon$-$\delta_{\rm m}$ plane the particle energy maps out different triangular regions, depending on the history of the excitation. 
The color scale indicates the effective temperature/oscillation amplitude.
To increase the contrast the top left map is scaled by a factor 2.
The upper row shows frequency sweeps at fixed modulation depth and the bottom row shows modulation depth sweeps at fixed modulation frequency.
The black dashed line marks the (linear) instability threshold \eqref{eq:Lock-inStabilityCondition}, which marks the transition from the low amplitude solution to the high amplitude solution
and the white dashed line marks the (nonlinear) instability threshold \eqref{eq:Lock-inStabilityConditionNL}, which marks the transition from the high amplitude solution to the low amplitude solution.
(b) Frequency up and down sweep at $\epsilon=22\times 10^{-3}$ (black dotted line in subfigure a top row).
The blue dashed line is a fit to \eqref{eqn:ao2}.
(c) Modulation depth up and down sweep at $\Omega_0\left/2\pi\right.=248.25\rm kHz$ and  (d) at $\Omega_0\left/2\pi\right.=250.25\rm kHz$  (black dotted lines in subfigure a bottom row).
The blue dashed line is a fit to \eqref{eqn:ao2}.
By fitting to the theoretical model \eqref{eqn:ao2} we extract the nonlinear parameters $\eta=6\rm  \mu m^{-2}$ and $\xi=-10\rm  \mu m^{-2}$.
\label{fig:FrequencyModulation}}
\end{figure*}
The data shown in Fig.~\ref{fig:FrequencyModulation} characterize the particle's response to an external parametric modulation close to the resonance frequency of the $x$-mode ($\Omega_x/2\pi\sim 125\rm kHz$).
Fig.~\ref{fig:FrequencyModulation}a shows maps of the particle amplitude in the $\epsilon$-$\delta_{\rm m}$-plane.
The boundaries, which are marked as dashed lines, are defined by the linear (black, Eq.~\ref{eq:Lock-inStabilityCondition}) and nonlinear (white, Eq.~\ref{eq:Lock-inStabilityConditionNL}) stability condition, respectively.
Positive and negative frequency scans (Fig.~\ref{fig:FrequencyModulation}a top) and modulation depth scans (Fig.~\ref{fig:FrequencyModulation}a bottom) carry the oscillator into the bistability region along different stable attractors.
This leads to hysteresis for the modulation parameters $\epsilon$ and $\delta_{\rm m}$ that fulfil both conditions \eqref{eq:Lock-inStabilityCondition} and \eqref{eq:Lock-inStabilityConditionNL}.
Fig.~\ref{fig:FrequencyModulation}b shows a frequency up and down sweep across the resonance for fixed modulation depth $\epsilon=22\times 10^{-3}$ (c.f. Fig.~\ref{fig:FrequencyModulation}a horizontal dotted line).
When the modulation frequency is increased, the particle amplitude follows the low amplitude solution while condition \eqref{eq:Lock-inStabilityCondition} is fulfilled.
Conversely, when the frequency is lowered the particle follows the high amplitude solution while condition \eqref{eq:Lock-inStabilityConditionNL} is true.
We also observe bistability for up and down sweeps of the modulation depth when the modulation frequency is less than twice the particle resonance frequency (c.f. Fig.~\ref{fig:FrequencyModulation}c).
If the modulation frequency is larger, the transition is smooth and does not exhibit hysteresis (c.f. Fig.~\ref{fig:FrequencyModulation}d).
Fitting the data to the theoretical prediction allows us to extract the nonlinear coefficients $\eta=6\rm  \mu m^{-2}$ and $\xi=-10\rm  \mu m^{-2}$.\\[-1ex]

Until now, we have neglected coupling between the three spatial modes.
However, for large oscillation amplitudes, the modes couple.
The coupling has the same origin as the cubic nonlinearities $\xi$.
For a Gaussian intensity distribution at the focus we find that the trap stiffness along $x$ is of the form \cite{Gieseler:2013ip}
\begin{equation}\label{eq:NonlinearCoupling}
k_{\rm trap}^{\!(x)}=m\Omega_i^2\left(1+\sum_{i=x,y,z}\xi^{\!(x)}_i x_i^2\right)
\end{equation}
and respective expressions hold for the trap stiffness along $y$ and $z$.
Here, $\xi^{\!(x)}_x=\xi$ is the Duffing nonlinearity we have considered so far and $\xi^{\!(x)}_y$ and $\xi^{\!(x)}_z$ are the nonlinear coupling coefficients of $y$ and $z$, respectively.
The nonlinear coupling gives rise to frequency shifts (pulling) of the orthogonal modes.
Eq.~\ref{eq:NonlinearCoupling} states that an increase in energy of any mode downshifts the frequency of the x-mode (note that $\xi^{(x)}_i$ is negative).
\begin{figure*}[th!]
 \begin{center}
\includegraphics[width=0.9\textwidth]{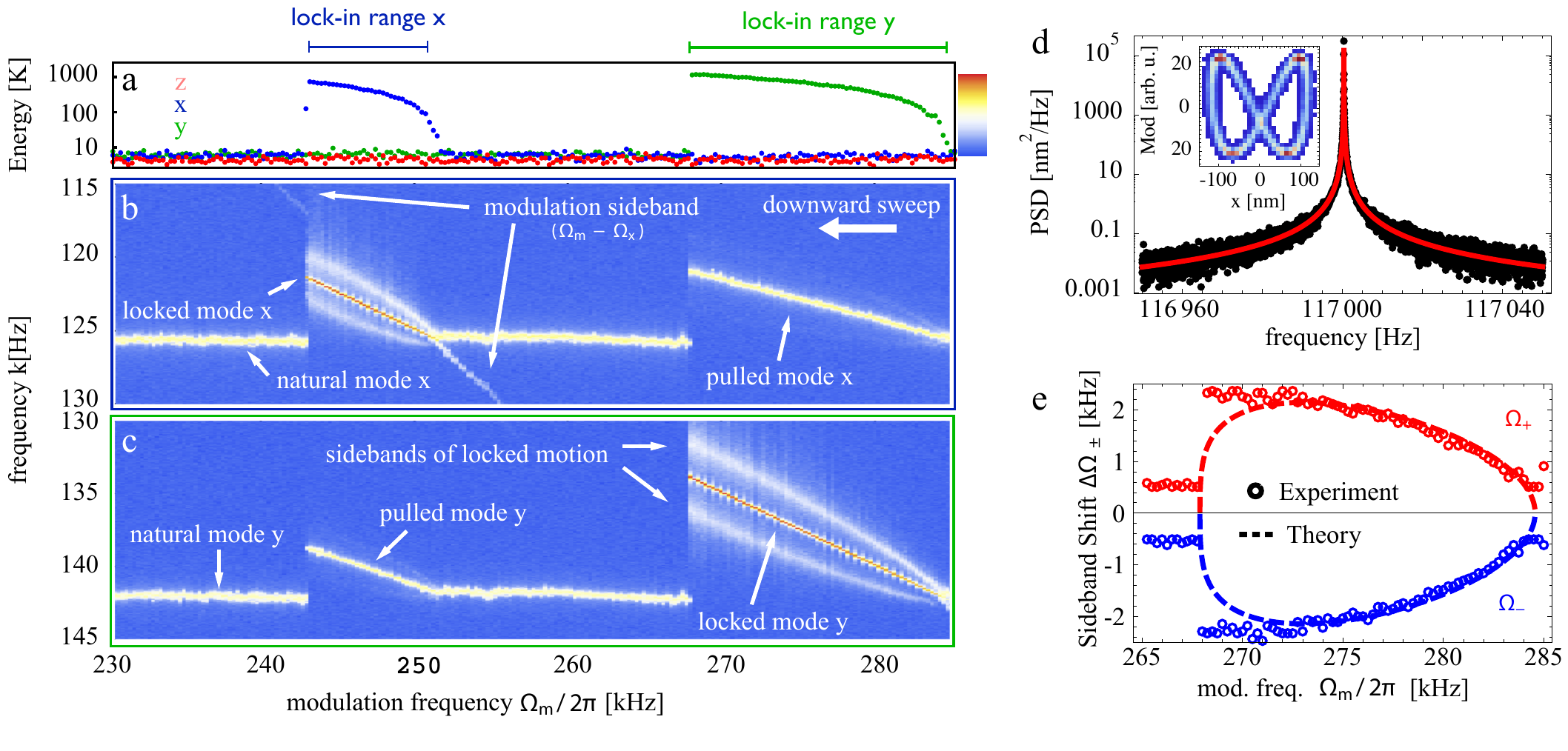}
\end{center}
\caption{
Nonlinear mode coupling and synchronization.
(a-c) The parametric modulation frequency $\Omega_{\rm m}/2\pi$ is reduced from 285kHz down to 230 kHz exciting the resonances of the $x$ and $y$ modes, respectively.
(a) Within the lock-in range energy is transferred to the resonant mode.
(b-c) The spectral map of the $i$th-mode ($i=x,y$) reveals that within the lock-in range the modulation sideband ($\Omega_{\rm m}-\Omega_i$) and the natural mode ($\Omega_i$) merge and give rise to a single oscillation at $\Omega_{\rm m}/2$.
If the  mode ($i=x(y)$) is driven resonantly, nonlinear coupling pulls the frequency of the off-resonant ($i=y(x)$) mode toward lower frequencies.
(d) The resonant mode phase locks to the external modulation.
The resonance is considerably sharper than the natural linewidth $\Gamma_0\sim 5\rm kHz$ defined by the vacuum pressure.
The observation of a Lissajous figure  in the density histogram of the particle motion versus the external modulation signal indicates that the particle phase is locked to the external modulation ( inset).
(e) Sidebands of $y$-mode as a function of modulation frequency $\Omega_{\rm m}$ extracted from subfigure c.
Open circles are the experimental data and dashed lines are the theoretical curves \eqref{eqn:SidebandFrequencies} with nonlinear coefficients $\xi=- 9\rm \mu m^{-2}$ and $\eta= 0.6 \rm \mu m^{-2}$ obtained from a fit to the amplitude response (c.f.~Fig.~\ref{fig:Spectra}a).
In Figs. a,b,c,e the pressure and modulation depth  are $1.6\times 10^{-4} \rm mBar$ ($Q\sim 10^6$) and $\epsilon=8\times 10^{-3}$, respectively. In Fig. d it is  $5.4\times 10^{-3} \rm mBar$ and $\epsilon=21\times 10^{-3}$.
\label{fig:Spectra}}
\end{figure*}
To illustrate the nonlinear coupling, we sweep the modulation frequency over a wide range covering both the $x$ and the $y$ resonances and record the power spectral density for each value of the modulation frequency.\\[-1ex]

Figs.~\ref{fig:Spectra}b,c show the resulting maps of the frequency components of the motion along $x$ and $y$, respectively, for a downward sweep at $1.6\times 10^{-4}\rm mBar$ with modulation depth $\epsilon= 8\times 10^{-3}$.
The parametric modulation frequency is reduced from $285\,\rm kHz$ down to $230\,\rm kHz$, exciting first the $y$-mode and then the $x$-mode.
Fig.~\ref{fig:Spectra}a shows that within the lock-in range given by \eqref{eq:Lock-inStabilityConditionNL}, energy is transferred to the resonant mode while the off-resonant modes stay at low energy.
Modulation at $\Omega_{\rm m}$ generates motional sidebands at $\Omega_i\pm\Omega_{\rm m}$.
When the lower sideband $\Omega_{\rm m}-\Omega_i$ approaches the resonance, it resonantly transfers energy to the particle and the energy of the resonant mode increases.
The modulation sideband is clearly visible in Fig.~\ref{fig:Spectra}b.
Starting at $256\rm kHz$ it approaches $2\Omega_x\sim 252\rm kHz$.
For $\Omega_{\rm m}-\Omega_x\leq 2\Omega_x$, the sideband and the natural mode $\Omega_x$ disappear and a strong mode at $\Omega_{\rm m}/2$ appears.
This mode remains stable as long as \eqref{eq:Lock-inStabilityConditionNL} holds.
For smaller values of $\Omega_{\rm m}$, the mode at $\Omega_{\rm m}/2$ vanishes (here at $\Omega_{\rm m}/2\pi\approx 243\rm kHz$) and the natural mode $\Omega_x$ and the modulation sideband $\Omega_{\rm m}-\Omega_x$ reappear.
Additionally, we observe the frequency-pulling of the non-resonant modes predicted by \eqref{eq:NonlinearCoupling}.
Note that the nonlinear frequency shifts as much as $5\rm kHz$, which corresponds to $\sim 10^7$ times the linewidth $\Gamma_0$!\\[-1ex]

Within the lock-in region the particle motion is phase locked to the external modulation as shown in Fig.~\ref{fig:Spectra}d.
Clearly, the resonance is much sharper than the expected $\Gamma_0\sim 5 \rm Hz$ at $5.4\times 10^{-3}\rm mBar$, indicating that the particle faithfully follows the modulation of the external source (Agilent 33521A).
Besides, since the phase between the modulation and the particle is fixed, we observe a butterfly shaped Lissajous figure when plotting the particle position against the modulation signal (c.f. inset Fig.~\ref{fig:Spectra}d).\\[-1ex]

Near the strong peak at $\Omega_m/2$ within the lock-in region, we observe sidebands (c.f. Figs.~\ref{fig:Spectra} b,c).
The sidebands originate from small perturbations of the steady state caused by collisions with residual air molecules.
Linearizing the equation of motion \eqref{eqn:EoMforA2} around the steady state \eqref{eqn:ao2} we find the characteristic frequencies
\begin{equation}\label{eqn:SidebandFrequencies}
  \Omega_\pm=\Omega_{\rm m}/2
  \pm
  \Omega_i\sqrt{
      \frac{3}{4}\xi q^2\left(\frac{3}{4}\xi q^2+\delta_{\rm m}\right)
      -\left(\frac{Q^{-1}}{2}\right)^2
      }
\end{equation}
of the sidebands.
Fig.~\ref{fig:Spectra}e shows the sideband shift $\Delta\Omega_\pm=\Omega_\pm-\Omega_{\rm m}/2$ extracted from Fig.~\ref{fig:Spectra}c to be in good agreement with the theoretical prediction \eqref{eqn:SidebandFrequencies}.\\[-1ex]
In conclusion, we have parametrically excited an optically levitated nanoparticle in high vacuum well into the nonlinear regime.
We have shown that each of the three individual spatial modes can be excited independently if parametric feedback keeps the oscillation amplitude below the thermal amplitude. For larger oscillation amplitudes
we observe nonlinear coupling between the modes. Using parametric coupling, the particle can be synchronized to an
 external source. A synchronized nanoparticle can act as a coherent nanoscale source of electric, magnetic \cite{Geiselmann:2013gb,Neukirch:2013ua} or gravitational forces \cite{Arvanitaki:2013ei}.
The present work  adds to our understanding of the dynamics of levitated nanoparticles in high vacuum and paves the way for applications in sensing \cite{Moser:2013go,Gieseler:2013ip} and macroscopic quantum mechanics \cite{Poot:2012fh,Aspelmeyer:2012to}.
Furthermore, nonlinear coupling can be explored for multimode sensing \cite{Hanay:2012ge}, phonon-cavity cooling \cite{Mahboob:2012cl} and frequency stabilization \cite{Antonio:1fj}. We also expect interesting results from further investigating the nanoparticle's  three-dimensional dynamics
in response to multi-frequency driving fields \cite{Westra:2010bn}, thermal excitation \cite{Gieseler:2013ip}, and coupling to other levitated nanoparticles \cite{Holmes:2012ew,Lechner:2013gz}.
This includes pattern formation \cite{Kenig:2009ii}, chaos \cite{Karabalin:2009dt} and stochastic resonance \cite{Gammaitoni:1900zz,Badzey:2005hl,Westra:2013du}.
Finally, in contrast to conventional nanomechanical oscillators, a levitated particle can rotate freely \citep{Arita:1ge}, thereby adding to the richness of the dynamics \cite{Manjavacas:2010bh}.


\end{document}